\documentclass[sigconf,nonacm]{acmart}
\usepackage{cleveref}

\usepackage{amssymb} % for checkmark and xmark symbols

\usepackage{xcolor}  % for colors
\usepackage{pifont}
\usepackage{multirow}
\usepackage{tabularx}
\usepackage{array}
\usepackage{footmisc}
\usepackage[most]{tcolorbox}
\newcommand{\cmark}{\textcolor{green!60!black}\checkmark}%
\newcommand{\xmark}{\textcolor{red}{\ding{55}}}%
\usepackage{adjustbox} % for centering the checkmark

\newcommand{\yellowcheck}{%
\fcolorbox{black}{yellow}{%
    \rule{0pt}{0.10em}\rule[0.10em]{0.10em}{0pt}\checkmark%
  }%
}
\AtBeginDocument{%
  }

\setcopyright{acmlicensed}
\copyrightyear{2018}
\acmYear{2018}
\acmDOI{XXXXXXX.XXXXXXX}
\acmConference[Conference acronym 'XX]{Make sure to enter the correct
  conference title from your rights confirmation email}{June 03--05,
  2018}{Woodstock, NY}
\acmISBN{978-1-4503-XXXX-X/2018/06}

\begin{document}

%%
%% The "title" command has an optional parameter,
%% allowing the author to define a "short title" to be used in page headers.
\title{Exploring Distributed Vector Databases Performance on HPC Platforms: A Study with Qdrant}

\settopmatter{authorsperrow=4}
\author{Seth Ockerman}
\email{sockerman@cs.wisc.edu}
\affiliation{%
  \institution{University of Wisconsin-Madison}
  \city{Madison}
  \state{Wisconsin}
  \country{USA}
}
\additionalaffiliation{
\institution{Argonne National Laboratory}
  \city{Lemont}
  \state{Ilinois}
  \country{USA}
}

\author{Amal Gueroudji}
\email{agueroudji@anl.gov}
\affiliation{%
  \institution{Argonne National Laboratory}
  \city{Lemont}
  \state{Illinois}
  \country{USA}
}
\author{Song Young Oh}
\email{so27@uchicago.edu}
\affiliation{%
  \institution{University of Chicago}
  \city{Chicago}
  \state{Illinois}
  \country{USA}
}
\author{Robert Underwood}
\email{runderwood@anl.gov}
\orcid{0000-0002-1464-729X}
\affiliation{%
  \institution{Argonne National Laboratory}
  \city{Lemont}
  \state{Illinois}
  \country{USA}
}
\author{Nicholas Chia}
\email{chia@anl.gov}
\affiliation{%
  \institution{Argonne National Laboratory}
  \city{Lemont}
  \state{Illinois}
  \country{USA}
}
\author{Kyle Chard}
\email{chard@uchicago.edu}
\affiliation{%
  \institution{University of Chicago}
  \city{Chicago}
  \state{Illinois}
  \country{USA}
}
\author{Robert Ross}
\email{rross@anl.gov}
\affiliation{%
  \institution{Argonne National Laboratory}
  \city{Lemont}
  \state{Illinois}
  \country{USA}
}
\author{Shivaram Venkataraman}
\email{shivaram@cs.wisc.edu}
\affiliation{%
  \institution{University of Wisconsin-Madison}
  \city{Madison}
  \state{Wisconsin}
  \country{USA}
}

%%
%% The "author" command and its associated commands are used to define
%% the authors and their affiliations.
%% Of note is the shared affiliation of the first two authors, and the
%% "authornote" and "authornotemark" commands
%% used to denote shared contribution to the research.
% \authornote{Both authors contributed equally to this research.}
% \email{trovato@corporation.com}
% \orcid{1234-5678-9012}
% \author{G.K.M. Tobin}
% \authornotemark[1]
% \email{webmaster@marysville-ohio.com}

%%
%% By default, the full list of authors will be used in the page
%% headers. Often, this list is too long, and will overlap
%% other information printed in the page headers. This command allows
%% the author to define a more concise list
%% of authors' names for this purpose.
\renewcommand{\shortauthors}{Ockerman et al.}

% Does the argumgent hold up

\begin{abstract}
Vector databases have rapidly grown in popularity, enabling efficient similarity search over data such as text, images, and video. They now play a central role in modern AI workflows, aiding large language models by grounding model outputs in external literature through retrieval-augmented generation. Despite their importance, little is known about the performance characteristics of vector databases in high-performance computing (HPC) systems that drive large-scale science. This work presents an empirical study of distributed vector database performance on the  Polaris supercomputer in the Argonne Leadership Computing Facility. We construct a realistic biological-text workload from BV-BRC and generate embeddings from the peS2o corpus using Qwen3-Embedding-4B. We select Qdrant to evaluate insertion, index construction, and query latency with up to 32 workers. Informed by practical lessons from our experience, this work takes a first step toward characterizing vector database performance on HPC platforms to guide future research and optimization. \footnote{To appear in the SC'25 Workshop Frontiers in Generative AI for HPC Science and Engineering: Foundations, Challenges, and Opportunities.}
\end{abstract}

\maketitle

\section{Introduction}
Vector databases enable efficient search over encoded representations of  embedded data known as vectors. Amid the rapid advancement of modern AI systems, they have become an integral component of scientific workflows~\cite{Kiran2025HybridRetrival, pan2023surveyvectordatabasemanagement, zhao2024fragfederatedvectordatabase}, particularly those leveraging retrieval-augmented generation (RAG)~\cite{barron2024domainspecificretrievalaugmentedgenerationusing, sarmah2024hybridragintegratingknowledgegraphs, Yang2025Dual, borgeaud2022improvinglanguagemodelsretrieving, fan2024surveyragmeetingllms}. As large-scale workflows are increasingly executed on high-performance computing (HPC) systems, vector databases must be adapted to the unique characteristics of these environments, which include specialized interconnects, parallel file systems, deep memory hierarchies, and heterogeneous hardware architectures~\cite{Latham2025Initial, Kwack2025AIandHPC, HomerdingPolarisAA, Lange2023_CloudHPC, sochat2025usabilityevaluationcloudhpc, Munhoz2023APerformance}. While prior work has studied the performance and trade-offs of vector databases~\cite{shen2024understandingsystemstradeoffsretrievalaugmented} in the context of single-GPU RAG, to the best of our knowledge no studies have focused on understanding or optimizing vector database performance in the context of scientific workloads and HPC systems, which remain the primary environment for large-scale scientific computation. A deeper understanding of how distributed vector databases perform on HPC architectures is necessary to inform system design, improve performance, and guide future research.

This work presents an early evaluation of vector database performance on an HPC system; we characterize the runtime performance of Qdrant \cite{qdrant2025}, a popular distributed vector database, on the Polaris supercomputer in the Argonne Leadership Computing Facility \footnote{\url{https://www.alcf.anl.gov/polaris}} using a realistic biological workflow. We generate embeddings based on the pes2o~\cite{peS2o} scientific text corpus using Qwen3-Embedding-4B~\cite{qwen3embedding}. We provide insight and recommendations for future work from our deployment experience on Polaris (see \cref{sec:conc}). In summary, we make the following contributions.

% , discovering that inference dominates the embedding generation runtime, Qdrant's data insertion time is a potential bottleneck for future HPC workloads, and that query workloads on modest datasets may not benefit from increased parallelization due to increased communication overhead.  In summary, we provide the following contributions:

\begin{itemize}
    \item We evaluate Qdrant's distributed performance on Polaris, testing insertion, index-building, and query performance with up to 32 Qdrant workers that span 8 compute nodes.
    \item We provide a first step toward characterizing vector database performance on HPC platforms, detailing the lessons learned from our experience.
    \item We publish a scientific embedding dataset and query workload for future use.\footnote{\url{https://doi.org/10.5281/zenodo.17101276}}
\end{itemize}

\section{Distributed Vector Databases}
\label{sec:survey}
\Cref{sec:background} provides the necessary background to understand the distributed vector database landscape. \Cref{sec:state} discusses a few of the popular distributed vector databases and their features. 

\subsection{Background}
\label{sec:background}
Vector databases are specialized data management systems designed to store, index, and search high-dimensional vector representations of data~\cite{johnson2017billionscalesimilaritysearchgpus, malkov2018efficientrobustapproximatenearest}. These vectors, also known as embeddings~\cite{mikolov2013efficientestimationwordrepresentations, pennington-etal-2014-glove}, are numerical representations of data such as text, images, or audio. Embeddings capture semantic or structural relationships between data such that similar items are represented by vectors that are close together in the embedding space~\cite{qwen3embedding, lee2025nvembedimprovedtechniquestraining}. This process enables efficient similarity search via (approximate) nearest neighbor search~\cite{Muja2014ScalableNN, Gionis1999Similarity}:  Given a query encoded as a vector, the system computes its distance (e.g., cosine similarity, euclidean, inner product) to all stored embeddings and returns the top $N$ closest vectors as the most similar results.

\begin{table*}[t]
\scalebox{0.90}{%
\begin{tabular}{|c|c|c|c|c|c|c|}
\hline
\textbf{System} & \textbf{Parallel Read/Write} & \textbf{Compute/Storage Separation} & \textbf{Load Balanced} & \textbf{Autoscaling} & \textbf{GPU Indexing} & \textbf{GPU ANN} \\
\hline
Vespa     & \cmark & \cmark & \yellowcheck &  \yellowcheck &  \xmark  & \xmark  \\ 
\hline
Vald      & \cmark & \xmark & \cmark &  \cmark & \cmark & \cmark \\
\hline
Weaviate  & \cmark &  \xmark & \cmark &  \cmark & \cmark & \cmark \\
\hline
Qdrant   & \cmark  & \xmark & \yellowcheck &  \cmark & \cmark  & \xmark   \\
\hline
Milvus    & \cmark & \cmark & \cmark &  \cmark & \cmark & \cmark \\
\hline
\end{tabular} %
}
\centering
\vspace{5pt}
\caption{Comparison of features among state-of-the-art distributed vector databases. Some of the listed features are  available only in the paid cloud offerings of the respective vector database; such entries are denoted as \yellowcheck. }
\label{tab:vectorFeatures}
\end{table*}

% In the context of vector search, recall measures the proportion of true nearest neighbors that are retrieved by the ANN and is formally defined in \cref{eq:recall}, where $R_\text{true}$ is the set of true nearest neighbors and $R_\text{retrieved}$ is the set of neighbors returned from the ANN search.
As the number of embeddings grows, searching the entire database becomes intractable~\cite{Muja2014ScalableNN}. To address this challenge, vector databases employ specialized data structures known as indexes~\cite{malkov2018efficientrobustapproximatenearest,Bentley1975Multidimensional,johnson2017billionscalesimilaritysearchgpus} to enable efficient approximate nearest neighbor (ANN) search. These indexes reduce the number of required distance computations by pruning large portions of the search space while aiming to maximize accuracy. Common index types include graph-based approaches such as Hierarchical Navigable Small World (HNSW) graphs~\cite{malkov2018efficientrobustapproximatenearest}, inverted file structures often paired with product quantization~\cite{Jegou2011Product}, and tree-based methods such as KD-trees~\cite{Bentley1975Multidimensional}. The choice of index depends on dataset size, dimensionality, latency requirements, and the desired trade-off between accuracy and query or insertion time. For details on algorithms and trade-offs, we refer readers to ~\citet{ma2025comprehensivesurveyvectordatabase}.

% \begin{equation}
% \label{eq:recall}
% \text{Recall} = \frac{|R_{\text{true}} \cap R_{\text{retrieved}}|}{|R_{\text{true}}|}
% \end{equation}

\begin{figure}[t]
 \centering
  \includegraphics[width=\columnwidth]{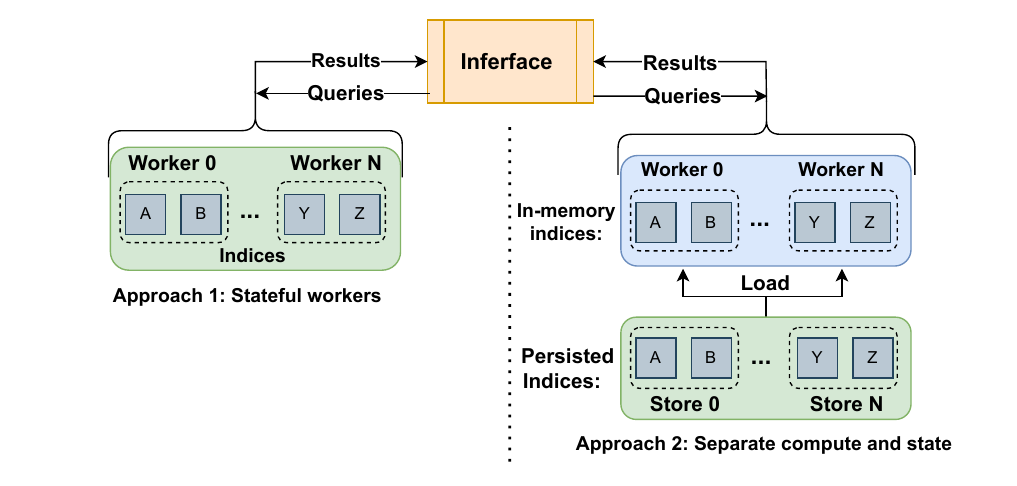}
% \vspace{-0.15in}
\caption{Two example distributed vector database configurations. Blue boxes represent stateless workers, and green boxes denote the presence of a state. }
\label{fig:twoDist}
\vspace{-10pt}
\end{figure}

% \footnotetext{\label{myfoot} Note that Vald supports GPU acceleration by providing compatibility with Faiss agents rather than providing native support.}

To achieve even greater scalability and support thousands of concurrent queries, practitioners employ distributed vector databases~\cite{Wang2021Milvus,vespa2025, qdrant2025,vald2025, weaviate2025}. Distributed vector databases divide coordination, computation, and data storage among multiple workers while presenting a single unified interface to users. In order to accomplish this, the data is sharded into independent indexes built for each shard~\cite{Wang2021Milvus,guo2022manucloudnativevector,vald2025,weaviate2025,qdrant2025,vespa2025}. Sharding is one of the primary techniques for achieving horizontal scalability in vector databases. There are two dominant sharding approaches: stateful (approach 1 of \cref{fig:twoDist}) and stateless with compute/storage separation (approach 2 of \cref{fig:twoDist}). In a stateful architecture, each worker stores state such as indexes or data and performs the needed computation to serve queries for its shard. In essence, the worker both ``owns" and is responsible for a portion of the dataset. This paradigm is used by vector databases such as Qdrant~\cite{qdrant2025}, Vald~\cite{vald2025}, and Weaviate~\cite{weaviate2025}. Alternatively, in a stateless architecture, workers perform computation but do not persistently store the dataset or indexes locally. Instead, data is stored in a separate, durable storage layer (often an object storage or file system) and loaded into a cache layer when needed. This approach is used by distributed vector databases such as Vespa~\cite{vespa2025} and Milvus~\cite{Wang2021Milvus}. Regardless of the specific architecture, a distributed vector database must support search across all data shards. To do so, the query is broadcast to all workers, \footnote{In the case of queries that filter based on a condition (predicated queries), some vector databases perform prefiltering to reduce the shard search space.  To the best of our knowledge, however, for non-predicated ANN search, all the systems discussed in this work follow a broadcast–reduce workflow.} and each worker performs an ANN search over its shards. The partial results are then aggregated, and the top results are returned.

% is this a reasonable arugment 
\subsection{State of the Art}
\label{sec:state}
A few popular distributed vector databases include Vespa~\cite{vespa2025}, Vald~\cite{vald2025}, Weaviate~\cite{weaviate2025}, Milvus~\cite{Wang2021Milvus,guo2022manucloudnativevector}, and Qdrant~\cite{qdrant2025}. \Cref{tab:vectorFeatures} shows an overview of a few of their notable features.  All the listed databases support parallel reading/writing, multicore acceleration, elasticity, and shard replication for increased availability and reliability. However, only a subset---Vespa and Milvus---support compute-storage separation, while only Vald, Weaviate, and Milvus support GPU-accelerated ANN search. The ability to scale compute independently of state allows the workflow to add more workers without repartitioning persisted data---traditionally an expensive process~\cite{Taipallus2024VectorDatabase, mohoney2025quakeadaptiveindexingvector, cheng2024characterizingdilemmaperformanceindex} that requires both data transfer and the reconstruction of impacted indexes. The degree to which compute–storage separation is critical depends on the workload. While all  the described vector databases support elastic addition/subtraction of workers, stateful architectures require data rebalancing before the new resources can be fully utilized. For relatively static query and update patterns, there is little need to rapidly scale the number of workers independently of data storage. However, recent work~\cite{mohoney2025quakeadaptiveindexingvector} showed that real-world workloads (e.g., Wikipedia) often exhibit dynamic and skewed access/update patterns, highlighting the advantages of compute-storage separation.

\subsection{Related Work}
% \begin{itemize}
% \item other survey works on Vector DBs
% \item any other works that eval Vector DBs, particularly if distributed
% \end{itemize}
The rapid adoption of vector databases in large language model (LLM) workflows and other data-intensive applications has led to several recent surveys~\cite{taipalus2024vector, pan2024vector, han2023comprehensive, kukreja2023vector, jing2025large} that review LLM architectures, storage/retrieval mechanisms, use cases, and open challenges. Although these works include feature-level comparisons of widely used systems, none provides empirical performance evaluations, particularly in HPC settings~\cite{oketunji2025high}. \citet{shen2024understandingsystemstradeoffsretrievalaugmented} evaluated multiple index types in the context of single-GPU RAG but did not evaluate distributed vector database systems or test in an HPC environment. \citet{xu2025harmony} proposed a distributed vector database designed for scalability and benchmarked it against FAISS ~\cite{douze2024faiss}, but they did not benchmark against existing distributed systems or perform experiments in an HPC setting.

\section{Performance Evaluation}
We consider an end-to-end workflow that leverages vector databases to contextualize raw data records with information from papers, which is intended to be used in biological RAGs.
This synthetic data could also be used in a variety of ways to improve LLM performance: pretraining/fine-tuning the model~\cite{gunasekar2023textbooksneed}, training a cross-modal adapter~\cite{NEURIPS2022_960a172b}, or better grounding the output of the system with tools (see \cite{openai_gpt-5_2025}). The target workload uses a small subset of 22,723 terms related to genomes available through BV-BRC ~\cite{BVBRC}---a comprehensive bioinformatics resource developed to support biological research. Each term is used to generate a query that searches the papers contained within the pes2o dataset~\cite{peS2o} (comprising up to 8 million full-text papers) for data related to the term. The intuition is that searching across a collection of research papers allows one to find data directly related to the target term, thereby providing better context for the information that would be supplied to a RAG system. This approach mimics prior work on synthetic data generation~\cite{gunasekar2023textbooksneed}.
Although pes2o is not a dedicated biological corpus, it serves as a proxy for an internal large corpus containing biological papers. In this work we focus on runtime performance rather than correctness, for which pes2o is sufficient. Our analysis examines embedding generation, data insertion, index-building, and query behavior. We perform all testing on Polaris. Each compute node features a 2.8~GHz AMD EPYC Milan 7543P 32-core CPU, 512~GB of DDR4 RAM, and four NVIDIA A100 GPUs. The system is interconnected using HPE Slingshot 11 and uses a Dragonfly topology. We select Qdrant as the vector database system for our initial evaluation.
%  \centering
%   \includegraphics[scale=0.45]{figures/embeddingFlow.drawio.png}
% \caption{Embedding generation pipeline. User-defined parameters control the batch size, queue priority, queue length, and GPUs per job.}
% \label{fig:embedding}
% \end{figure}

\subsection{Embedding Generation}
We generate embeddings using the collection of full academic papers in the pes2o dataset, comprising a total of 8,293,485 embeddings. We generate a single embedding per paper by feeding each paper's full text into the Qwen3-Embedding-4B model, a state-of-the-art embedding model that fits within a single 40~GB GPU. In future work we could apply chunking techniques~\cite{smith_evaluating_2024}, which would likely improve retrieval quality but increase the number of entities in the database, stressing performance further. To ensure efficiency, we design an adaptive pipeline overseen by an orchestrator. Based on user-controlled parameters, the orchestrator batches the input text into single-node jobs to minimize queue wait time and monitors a user-defined set of queues. As availability within a queue opens, the orchestrator submits the next batch. The orchestrator can be paused and resumed as needed, with the flexibility to adjust target queues and the number of jobs per queue. Within a single job, multiprocessing is used to process papers concurrently, splitting work among all available GPUs. Each GPU uses a simple heuristic---based on limits for total characters and the number of papers per batch---to determine how many papers to process in each batch. Based on empirical observations, we define each batch as 4,000 papers and set the total batch character limit and maximum batch size to 150,000 and 8, respectively. In the event of an OOM error, the GPU falls back to sequential processing for that individual batch, ensuring that there is no possibility of truncated papers. \\

\noindent \textbf{Results: } Across all jobs, embedding generation (model inference) dominates overall runtime (see \cref{tab:simple2row}), with a mean runtime that comprises 98.5\% of total runtime ($2,417.84 \pm 113.92$~s). Notably, the batching heuristic was highly successful at preventing memory errors while promoting parallelism, processing less than 0.10\% of the papers sequentially. \textbf{These findings indicate that for datasets that fit comfortably within an HPC compute node’s memory, embedding generation efforts should prioritize improving the efficiency of model inference rather than I/O or model loading.}

% Model loading and I/O represent a smaller portion of runtime relative to inference
% We observed 37 outliers in overall runtime as defined by the IQR and $3\sigma$ rules, which coincide with fluctuations in inference time. While model loading and I/O exhibit higher relative variability, their magnitudes are not sufficient to cause significant fluctuations.

% \begin{tcolorbox}[colback=blue!5!white, colframe=blue!75!black,
%   title=Lesson 1, fonttitle=\bfseries]
% Inference dominates embedding generation runtime, and heuristic batching effectively maximizes parallelism.
% \end{tcolorbox}

\begin{table}[t]
\centering
\begin{tabular}{|c|c|c|}
\hline
\textbf{Model Loading} & \textbf{I/O} & \textbf{Inference}  \\
\hline
  28.17 & 7.49 & 2381.97  \\
  % 28.17 $\pm$ 16.37  & 7.49 $\pm$ 4.34 & 2381.97 $\pm$ 113.10   \\
  % 28.17 $\pm$ 16.37  & 7.49 $\pm$ 4.34 & 2398.56 $\pm$ 113.10 & 2417.84 $\pm$ 113.92  \\
\hline
\end{tabular}
\vspace{5pt}
\caption{Mean embedding generation runtime in seconds across $N{=}2{,}079$ batches of approximately 4,000 papers. Model loading refers to loading the model weights from disk and transferring them to the GPU; I/O denotes the time spent loading the raw text from disk; and inference refers to the period spent generating embeddings.}
\label{tab:simple2row}
\vspace{-30pt}
\end{table}
% Runtime variance tracked inference, with 37 flagged outliers for inference and 39 for total time (detected using both , while model loading and I/O showed higher relative variability at smaller scales. 

% As shown in \autoref{tab-embedding}, GPU-level analysis revealed minor heterogeneity: GPU~2 had the lowest average total time (2398.43~s), GPU~3 the highest (2434.50~s) yet most consistent (std 49.43~s), and GPU~1 showed substantial noise (std 162.44~s). 

% The batching heuristic was highly successful at preventing memory errors while promoting parallelism, processing only TODO\% of the data sequentially. 

\subsection{Data Insertion}
\label{sec:dataUpload}
After embedding generation, the data must be uploaded to the Qdrant workers. To optimize insertion performance, we tune the batch size (i.e., number of vectors per upload request) and the number of allowed concurrent upload requests on a 1~GB subset of the full dataset. Although the effects of changing batch size and concurrency may interact, for brevity in this work we fix the batch size to the optimal value discovered during batch size tuning while adjusting the degree of concurrency. To perform multiple concurrent upload requests, we use Qdrant's asynchronous client implementation and Python's \texttt{asyncio} library. After tuning, we upload the full dataset to a Qdrant cluster with the following number of workers: 1, 4, 8, 16, and 32. The data is partitioned across workers, with each worker responsible for approximately $\text{80~GB}/\text{\#Workers}$ of data. We employ multiprocessing to assign one client to each Qdrant worker. Each client is configured with the optimal batch size and degree of concurrency determined during tuning. All clients run on a single compute node, while the Qdrant servers are deployed on separate compute nodes, with four Qdrant workers per machine.

% While higher throughput with a single client may be possible by running multiple parallel instances of the client via multiprocessing, we focus here on the default single-client asynchronous setup to better approximate the out-of-the-box baseline experience.
% To minimize node-hour consumption, we tested index build time within the same jobs (see \cref{sec:index} for more details) by pausing insertion at pre-defined amounts of total data to measure index build time; the time spent on index testing was not included in this section's reported measurements.

% multi-process synchronization across clients was required to ensure that insertions are paused at the correct stage. Consequently, the multi-worker configurations could potentially achieve faster performance in practice without the added synchronization mechanisms we needed for testing; however, the reported results provide a baseline for insertion performance. \\

% batched vs seqentual
% insertion time with increasing numbers of nodes
% index consturction time with more data / more nodes/ more shards

% index build time 
    % varied amount of data

% split data, insert as we go
    % 1. batch size
    % vary number of nodes
    % threshold

\begin{figure}[t]
 \centering
  \includegraphics[width=\columnwidth]{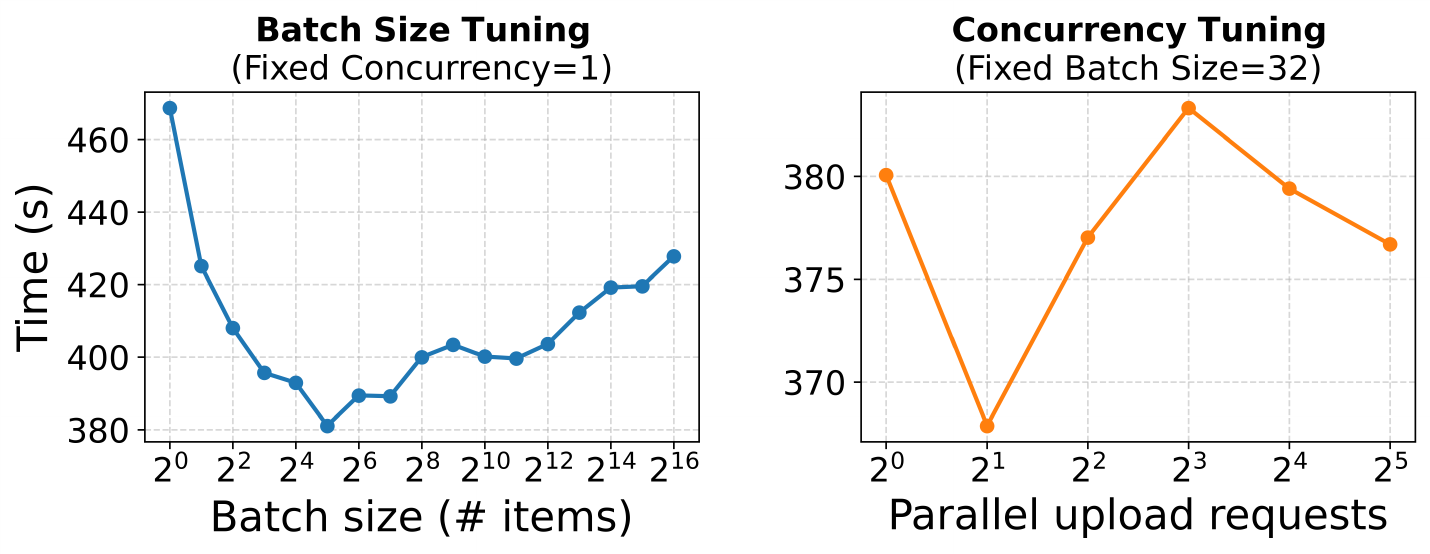}
\caption{Data insertion time for a 1~GB dataset into a single-worker Qdrant cluster on Polaris  using varying batch sizes and parallel requests. The optimal discovered batch size was used while tuning the number of parallel requests. }
\label{fig:insertion}
\vspace{-10pt}
\end{figure}

\noindent \textbf{Results: } \Cref{fig:insertion} presents the insertion time for a 1~GB subset of the full dataset, measured using a single Qdrant worker with varying parameter settings. Batch size exhibits a clear optimization curve, with performance improving from $468\,\mathrm{s}$ (size~1) to a minimum of $381\,\mathrm{s}$ (size~32) before gradually degrading at larger batch sizes. Increasing the number of concurrent insertion requests shows diminishing returns: insertion time decreases from $381\,\mathrm{s}$ (1~request) to $367\,\mathrm{s}$ (2~requests) but increases thereafter. This trend reflects the constraints of \texttt{asyncio} when applied directly to data insertion without further customization. By default Python's \texttt{asyncio} library runs tasks in a single synchronous thread, with each task  yielding control only when it hits the await keyword during data upload; CPU-bound tasks are not performed in parallel. Profiling reveals that, on average, with a batch size of 32, converting the batch into a Qdrant batch object---a CPU task---for upload requires 45.64 ms, while data insertion requires only 14.86 ms. Thus, the potential speedup from allowing multiple concurrent upload requests is minimal, defined at a maximum of 1.31$\times$ by Amdahl's law~\cite{Amdahl1967Law}. 
% \textbf{GAIL - having four lines of text in bold is not recommended. Either pick a few words to emphasize or delete the boldface.}
Qdrant's asynchronous approach to single-client parallelism yields limited speedup during data upload, as CPU-bound tasks dominate runtime. Consequently,\textbf{ multiprocessing may be better suited than \texttt{asyncio} for single-client parallelism during data insertion.} The scaling is more favorable as we increase the number of Qdrant workers and correspondingly total clients. The total insertion time decreases from approximately 8.22 hours with 1 Qdrant worker to 21.67 minutes with 32 Qdrant workers (see \cref{tab:qdrant_upload}). While the upload speed is significantly below the theoretical network bandwidth, this is expected; during data insertion, in addition to the data being communicated over the network, Qdrant is storing the data, optimizing the data layout to minimize memory usage, and building indexes in the background. 
% \textbf{GAIL - and here especially delete the bold except perhaps around "the rate...workloads".}
While a more detailed profile of I/O, data communication, and CPU operations is needed to understand the cause, the rate of \textbf{data insertion has the potential to become a bottleneck for large-scale, scientific HPC workloads} that need to continually insert, index, and search new data. Further optimizations to data insertion should be a high priority for the HPC community.

% \begin{tcolorbox}[colback=blue!5!white, colframe=blue!75!black,
%   title=Lesson 2, fonttitle=\bfseries]
% Qdrant's default asynchronous approach to single-client parallelism yields limited speedup during data upload, as CPU-bound tasks dominate runtime. Consequently, multiprocessing is preferred over \texttt{asyncio} for single-client parallelism during data upload.
% \end{tcolorbox}

% As noted in the documentation, while it is recommend to leave the memory optimizations in place during ingestion to reduce memory consumption, it is possible to disable them entirely to potentiall increase overall throughput. A follow up experiment measuring upload time of 1~GB of data  showed that the runtime was reduced from TODO to TODO by disabling memory-layout optimizations. \textbf{This suggests that in HPC environments, which typically feature sizable amounts of RAM, it is beneficial to reduce memory optimizations during data upload to increase the portion of utilized network throughput. }

% \begin{table}[t]
% \centering
% \begin{tabular}{|c|c|}
% \hline
% \textbf{\# Qdrant Workers} & \textbf{Upload Time} \\
% \hline
% 1 & 8.22 hours \\
% \hline
% 4 & 2.11 hours \\
% \hline
% 8 & 1.14 hours \\
% \hline
% 16 & 35.92 minutes \\
% \hline
% 32 & 21.67 minutes \\
% \hline
% \end{tabular}
% \vspace{5pt}
% \caption{Full dataset ($\approx$80~GB) upload time as a function of the number of Qdrant workers.}
% \label{tab:qdrant_upload}
% \vspace{-20pt}
% \end{table}

\begin{table}[t]
\centering
\begin{tabular}{|c|c|c|c|c|c|}
\hline
\textbf{Workers} &1 & 4 & 8 & 16 & 32 \\
\hline
\textbf{Time} & 8.22 h & 2.11 h & 1.14 h & 35.92 m & 21.67 m \\
\hline
\end{tabular}
\vspace{5pt}
\caption{Full dataset ($\approx$80~GB) insertion time as a function of the number of Qdrant workers.}
\label{tab:qdrant_upload}
\vspace{-20pt}
\end{table}

\subsection{Index-Building}
\label{sec:index}
To evaluate the index-building phase, we measure index construction time with various amounts of data. Although indexes are typically built incrementally as data arrives, Qdrant’s documentation\footnote{\url{https://qdrant.tech/documentation/database-tutorials/bulk-upload/}} suggests deferring index construction to accelerate insertion in certain cases, necessitating a complete index rebuild. We mimic this scenario and use the default HNSW index settings. For this work we focus on CPU evaluation; future work will explore Qdrant's performance with GPU-enabled index-building.\\

% By design, indexes are not built over data until the volume of unindexed data exceeds 20,000~KB, since exhaustive search is more efficient below this threshold.
% \begin{figure}[t]\\

\begin{figure}[t]
 \centering
  \includegraphics[width=\columnwidth]{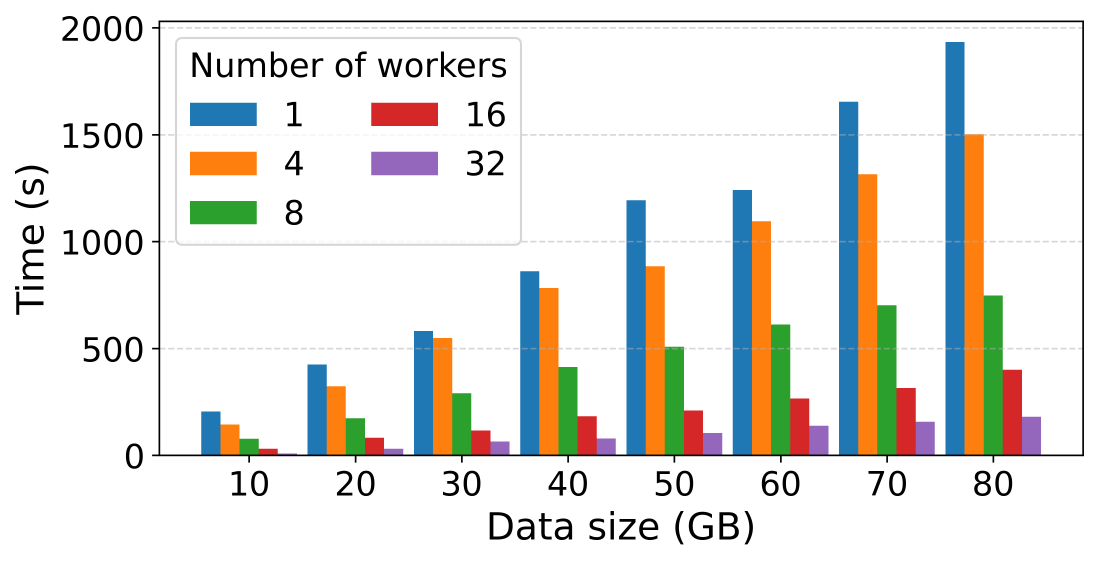}
\caption{Index build time versus dataset size for varying numbers of Qdrant workers.}
\label{fig:index}
\end{figure}

\noindent \textbf{Results:} As the number of Qdrant workers increases, the index build time decreases, with a maximum speedup of 21.32$\times$ using 32 workers relative to a single-worker Qdrant. This scaling behavior is expected because each index can be constructed independently; partitioning the data across workers proportionally reduces the workload per worker and enables substantial performance gains. However, as shown in \cref{fig:index}, the scaling falls short of linear. This is likely due to interworker communication overhead and resource contention, as each group of four workers shares a single compute node. This limitation is most apparent when scaling from one to four workers, which displays a maximum speedup of 1.27x. 
% \textbf{GAIL - perhaps limit bold to the part "deploying ..."}
Profiling reveals that a single worker already utilizes 90-97\% of the compute node's CPU capacity during index construction, indicating that \textbf{deploying multiple Qdrant workers per node is unnecessary to achieve CPU saturation during index-building.} To better exploit per-node resources and leverage multiple Qdrant workers per node, index-building could be offloaded to GPUs. Future work will test different cluster configurations and GPU-based index construction.

\subsection{Query}
To optimize query performance, we tune the query batch size and number of concurrent batches in flight in the same manner as described in \cref{sec:dataUpload}.  We set the query to return the 10 most similar results. After tuning, we test our biological query workload with Qdrant clusters of 1, 4, 8, 16, and 32 workers, utilizing the parameters discovered through tuning. \\ 

% query when all data avaiable
    % vary number of nodes

% load balancing
% 

% single node - increasing numbers of cores

% entire workflow
% increasing number of workers single node
% multi-node (up to 16)
% client patterns

\begin{figure}[t]
 \centering
  \includegraphics[width=\columnwidth]{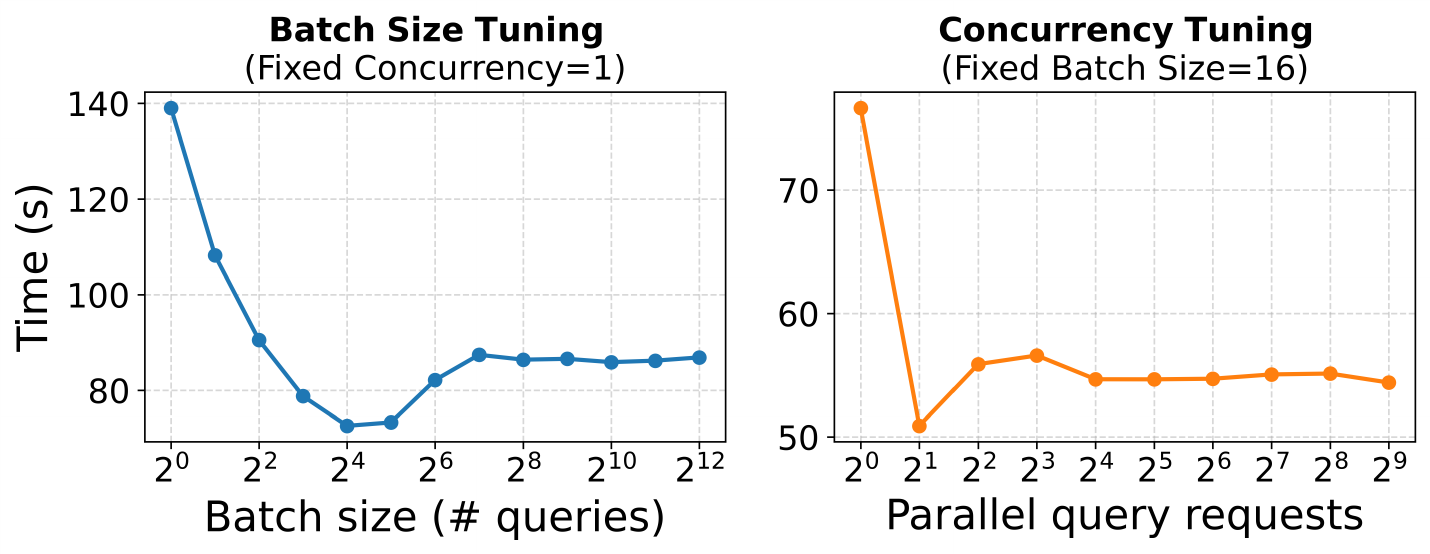}
\caption{Query running time for a 1\,GB dataset into a single-worker Qdrant cluster on Polaris  using varying batch sizes and parallel requests. }
\label{fig:query}
\vspace{-7pt}
\end{figure}

\noindent \textbf{Results: } \Cref{fig:query} shows query time using a single Qdrant worker with varying parameter settings on a 1~GB subset of the data. We observe that increasing the batch size reduces runtime until a batch size of 16 (from \(139\,\mathrm{s}\) to \(73\,\mathrm{s}\)) before further increases yield minimal benefit. Similar to the results shown in \cref{tab:qdrant_upload}, the shortest runtime is observed when only two parallel query requests are allowed. Follow-up testing revealed that as the number of parallel batch requests increases past 2, the average time spent waiting for the result from the worker grows correspondingly. For example, the average per-batch call time rises from 30.7 ms with 2 concurrent requests to 76.4 ms with 4 requests, and further to 170 ms with 8 requests, suggesting that the worker's resources are saturated. In our distributed tests, increasing the number of workers provides little benefit until the dataset reaches at least 30~GB (see \cref{fig:query}). This behavior arises from Qdrant’s query execution model: the client submits a query to one of the workers, which broadcasts it to the others. Each worker then searches its local shards and returns partial results to the worker first contacted by the client, which sends the final response back. Although this approach parallelizes the search computation, it also introduces communication overhead across the workers. For smaller datasets, this overhead outweighs the gains from horizontal sharding; only once the dataset size exceeds 30~GB does the parallelization begin to deliver a speedup, reducing runtime by a maximum of 3.57$\times$. Notably, increasing the cluster size beyond four provides only marginal improvements, suggesting that the reduction in runtime due to parallelization may be overshadowed by the cost of interworker communication. \textbf{Our results suggest that further improvement could be obtained if the cluster could adaptively scale based on the size of the data.}

% the for smaller datasets the additional communication overhead of multi-worker/multi-node approaches outweighs the benefits of further parallelization.}
% Profiling revealed two contributing factors: high RPC wait time and high CPU utilization on the worker nodes. 

% Increasing parallel query requests helps from 1 to 2 requests (\(76.6\,\mathrm{s}\,\rightarrow\,50.9\,\mathrm{s}\)) but then stays \(54\)--\(56\,\mathrm{s}\) up to parallel 512 requests.

\begin{figure}[t]
 \centering
  \includegraphics[width=\columnwidth]{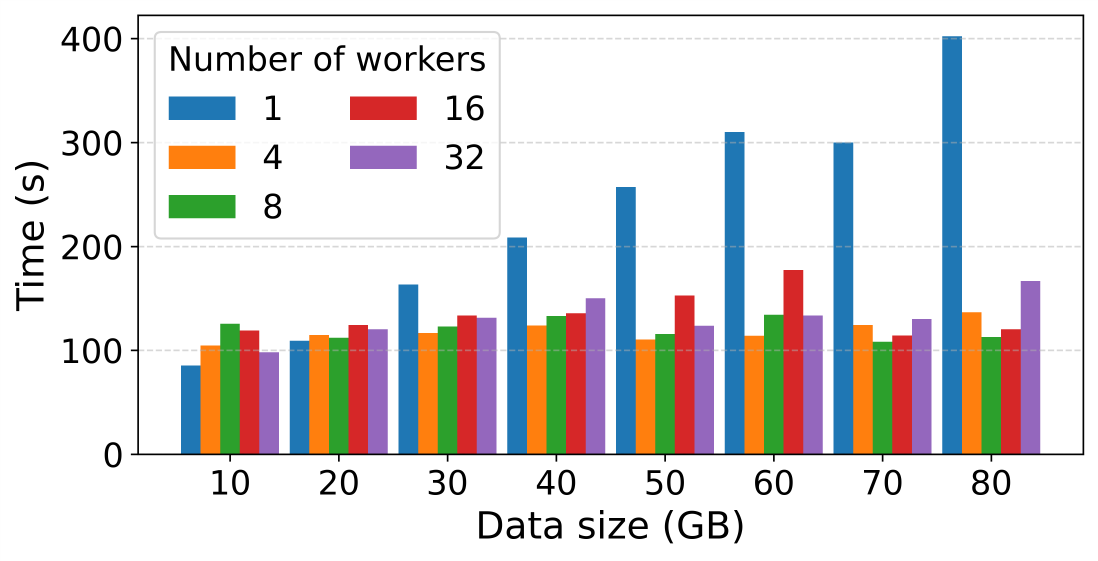}
\caption{Query time versus dataset size for varying numbers of Qdrant workers.}
\label{fig:query_time}
\vspace{-15pt}
\end{figure}

% Beyond surveys, other system evaluation efforts for Vector DBs remain limited. A prior study \cite{wang2025towards} examined bug patterns in open-source Vector DBs and proposed criteria for software quality assessment but did not extend their analysis to system-level performance. Another study \cite{wang2025towards2} introduced a new retrieval quality metric for Vector DBs that improves upon average recall, yet their experiments focused on comparing indexing algorithms (e.g., HNSW, ScaNN) rather than complete database systems. In the distributed setting---our primary focus---recent work \cite{xu2025harmony} proposed a system design focused on scalability and benchmarked it against FAISS ~\cite{douze2024faiss}, but without broader comparisons to other Vector DBs. 

% Our work addresses that gap by presenting the first comprehensive end-to-end evaluation of a distributed vectorDB on an HPC platform.

\section{Conclusion}
\label{sec:conc}
This work presents an initial evaluation of the distributed vector database system, Qdrant, in an HPC environment with up to 32 workers. We evaluate a realistic end-to-end biology workload, including embedding generation, data insertion, index-building, and query runtime. We release our embedding and query dataset for future use, and we provide the following initial insights based on our experience:
\begin{itemize}
    \item Embedding generation runtime is dominated by model inference.
    \item The conversion of data into Qdrant batch objects is CPU-bound and often slower than the insertion RPC, making multiprocessing a better choice than \texttt{asyncio}.
    \item Index-building is a CPU-intensive workload, saturating a compute node's CPU while utilizing only a single worker. Offloading index-building to the GPUs may increase the benefit of utilizing multiple workers per compute node.  
    \item Increasing the number of workers yielded only limited reductions in query runtime for our 80 GB dataset. Additional techniques may be required to fully leverage multiworker parallelism on smaller datasets.
\end{itemize}
In this study we did not focus on runtime variability or reproducibility. Future work could investigate the performance variability. We also evaluated only CPU-based index construction; a comparison against the GPU implementation is warranted in future work. Moreover, our evaluation focused on a single system; a comprehensive, multisystem study of distributed vector databases on different HPC platforms is needed to fully characterize the design space. \footnote{ChatGPT \cite{openai2024chatgpt} and Grammarly \cite{grammarly2024} were used to improve the grammar and phrasing of this work.}

\begin{acks}
This material is based upon work supported by Laboratory Directed Research and Development (LDRD) funding from Argonne National Laboratory, provided by the Director, Office of Science, of the U.S. Department of Energy under Contract No. DE-AC02-06CH11357.
An award of computer time was provided by the INCITE program. This research used resources of the Argonne Leadership Computing Facility, which is a DOE Office of Science User Facility supported under Contract DE-AC02-06CH11357.
\end{acks}

\bibliographystyle{ACM-Reference-Format}
\bibliography{main}

\end{document}